\documentclass[a4paper,showkeys,aps,prd,reprint,nofootinbib,showpacs,twocolumn]{revtex4-1}
\usepackage{hyperref,manfnt}
\usepackage{microtype}%				
\usepackage[tbtags]{amsmath}
\usepackage{mathrsfs}
\usepackage{amssymb}
\usepackage{amsthm}
\usepackage{amscd}
\usepackage{wasysym}
\usepackage{tensind}%				
\tensordelimiter{?}
\whenindex{;}{\,;\,}{\finishdots}
\whenindex{,}{\,,\,}{\finishdots}
\whenindex{|}{\,\vert\,}{\finishdots}
\usepackage[roman,text,thinc]{esdiff}%		
\usepackage{esint}%				
\usepackage{yhmath}%				
\usepackage{url}%					
\newcommand{\pd}{\partial}
\newcommand{\de}{\mathrm{d}}

\newcommand{\ton}[1]{\left(#1\right)}
\newcommand{\qua}[1]{\left[#1\right]}

\newcommand{\abs}[1]{\left\vert#1\right\vert}

\newcommand{\subeq}[1]{\begin{subequations} #1 \end{subequations}}
\newcommand{\eq}[1]{\( #1 \)}
\newcommand{\eqd}[1]{\begin{equation} #1 \end{equation}}

\newcommand{\speq}[1]{\begin{equation}\begin{split}#1\end{split}\end{equation}}

\newcommand{\gat}[1]{\begin{gather} #1 \end{gather}}

\newcommand{\bol}[1]{\mathbb{#1}}
\newcommand{\Cal}[1]{\mathcal{#1}}
\newcommand{\Scr}[1]{\mathscr{#1}}

\newcommand{\matg}{\text{\textsl{g}}}%	Upright slanted lowercase 'g', math mode
\theoremstyle{plain}
\newtheorem*{GJ}{Theorem (Geroch, Jang)}

\newtheorem*{NEP}{Newton's Equivalence Principle}
\newtheorem*{GWEP}{Gravitational Weak Equivalence Principle}
\begin{document}

\title{Weak equivalence principle for self-gravitating bodies: \\
A sieve for purely metric theories of gravity}

\date{January 2014}

\author{\firstname{Eolo} \surname{Di Casola}}
\email[Corresponding author. E-mail address: ]{eolo.dicasola@sissa.it}
\author{\firstname{Stefano} \surname{Liberati}}
\email[E-mail address: ]{stefano.liberati@sissa.it}
\affiliation{SISSA, via Bonomea 265, 34136 Trieste (Italy)}
\affiliation{INFN Sez. di Trieste, via Valerio 2, 34127 Trieste (Italy)}
\author{\firstname{Sebastiano} \surname{Sonego}}
\email[E-mail address: ]{sebastiano.sonego@uniud.it}
\affiliation{DCFA, Sezione di Fisica e Matematica, Universit{\`a} di Udine, Via delle Scienze 206, 33100 Udine (Italy)}

\begin{abstract}
We propose the almost-geodesic motion of self-gravitating test bodies as a possible selection rule among metric theories of gravity.  Starting from a heuristic statement, the ``gravitational weak equivalence principle'', we build a formal, operative test able to probe the validity of the principle for any metric theory of gravity, in an arbitrary number of spacetime dimensions.  We show that, if the theory admits a well-posed variational formulation, this test singles out only the purely metric theories of gravity.  This conclusion reproduces known results in the cases of general relativity (as well as with a cosmological constant term), and scalar-tensor theories, but extends also to debated or unknown scenarios, such as the $f (R)$ and Lanczos--Lovelock theories.  We thus provide new tools going beyond the standard methods, where the latter turn out to be inconclusive or inapplicable.
\end{abstract}


\maketitle

%===========================
\section{Introduction}
\label{S:1}
%===========================

Despite the pivotal role played at the dawn of general relativity, the ``principle of equivalence'' is now regarded as a rather vague, heuristic statement, perhaps useful for teaching purposes, but surely unable to challenge the contemporary approach to gravity, based on sharp and neat mathematical axioms and precision-test experiments~\cite{will, will2006lr}. Yet, the original idea behind the statement was nothing but brilliant: to provide a direct connection between the basic structural elements of a theory and  intuitive elementary  experiments~\cite{ess1}.

Is it possible to reconcile the two points of view, building a rigorous model out of a crisp, yet somewhat imperfect, premise? We believe so, and we try to prove it by showing, in the present work, how an aptly refined version of the principle of equivalence can be smoothly translated into formal statements without shadowing the underlying physical intuition. We then make use of our construction to establish a few selection criteria in the arena of extended theories of gravitation, singling out a specific sub-class of frameworks.

To this end, we must first and foremost notice that ``equivalence principle'' is way too generic an expression.  It is actually possible to distinguish several inequivalent formulations of the principle, each one identifying a different class of theories, with different fundamental mathematical structures.  These principles are sometimes formulated rather loosely, and different authors present them in ways that are not exactly equivalent.  There is, however, general agreement about the formal implications and physical content of some of them.\footnote{The review offered here of the various equivalence principles is inevitably limited and incomplete, as we highlight only those aspects that are relevant for the purpose of the present paper.  See Ref.~\cite{ess1} for a more detailed examination of the various elements entering into the formulation of each of these principles.}  Broadly speaking, the universality of free fall for test bodies with negligible self-gravity (\emph{weak equivalence principle}, or WEP) implies that gravity can be described in terms of a nonflat connection \eq{\Gamma} on  spacetime.

The eliminability of gravity in the local description of fundamental, nongravitational phenomena (\emph{Einstein's equivalence principle}, or EEP), together with the generally accepted Minkowskian character of spacetime in the absence of gravity, implies that \eq{\Gamma} is the Levi-Civita connection of a metric \eq{\matg_{ab}} with a Lorentzian signature, thus singling out the so-called \emph{metric theories of gravity}.\footnote{This is the very minimum requirement, and nothing forbids the presence of other gravitational degrees of freedom (scalars, vectors, tensors, etc.) in addition to the metric.} Moreover, within such theories, EEP provides a prescription for writing down physical laws in a gravitational field once they are known in Minkowski spacetime.

In the literature, one often finds yet another equivalence principle: the so-called \emph{strong equivalence principle} (SEP), which extends the validity of EEP to gravitational phenomena as well; i.e., it postulates that a background gravitational field does not influence the outcome of local experiments involving gravity.  The SEP implies of course EEP, but the converse is not true, so it is legitimate to ask which further restrictions does the SEP imposes upon a metric theory of gravity.

In this respect, it is useful to introduce also a somewhat intermediate step, in the form of a \emph{gravitational weak equivalence principle} (GWEP). The latter extends the universality of free fall to self-gravitating test bodies.\footnote{Those of being ``self-gravitating'' and ``test''  are logically and physically independent properties.  Whether a body has a sizeable self-gravity, is something which can be established  considering the ratio between its gravitational self-energy and its rest energy.  On the other hand, a test body is simply one not affecting appreciably its environment. This subtle, yet relevant difference is further examined in Ref.~\cite{ess1}.} The GWEP is clearly a necessary condition for the SEP (actually, its main founding pillar), but is not implied by EEP, as the latter deals exclusively with non-gravitational phenomena.  Moreover, the validity of the GWEP is in many ways easier to establish than the one of the SEP, being linked, as we shall see, to clearer and more familiar geometrical statements. For all these reasons, we deem it appropriate to focus first and foremost on the GWEP itself, its formal implications, and operational implementation.

In four spacetime dimensions, the GWEP turns out to be an extremely selective requirement.  Only the theories of Einstein and Nordstr{\"o}m abide by such principle.\footnote{The scalar theory of gravity proposed by Gunnar Nordstr{\"o}m  is a neglected gem of forgotten lore, which gave an impressive boost to the development of Einstein's grand synthesis.  See Refs.~\cite{Ravndal:2004ym, Giulini:2006ry} for a recollection of the genesis and features of this theory.}  It is therefore interesting to know what characterises these theories and makes them ``special''.  Also, it is desirable to export --- when possible --- the results to the vast landscape of extended theories of gravity, and to higher-dimensional spacetimes (where there is a wide range of ``natural extensions'' of general relativity, the so-called Lanczos--Lovelock theories).

We start by discussing the formulation of the GWEP in Sec.~\ref{S:2}.  Differently from what happens for the WEP, even in the case of a ``minimal'' theory like general relativity, it is not obvious at first sight whether the GWEP holds or not.  Thus, in Sec.~\ref{S:5}, we develop a method, encompassing a wide class of gravitational theories (and in an arbitrary number of spacetime dimensions), to check the validity of the GWEP for a given framework.  In Sec.~\ref{S:6}, the method is applied to the specific cases of general relativity (as well as in the presence of a cosmological constant), scalar-tensor and higher-order theories, and Lanczos--Lovelock/Gau\ss{}--Bonnet theories.  Our conclusions are presented in Sec.~\ref{S:7}.

Hereinafter, spacetime is modelled by a smooth \eq{m}-dimensional  Lorentzian manifold \eq{\ton{\Scr{M} , \matg_{ab}}}.  Physical fields are geometric objects (tensorial or spinorial) on \eq{\Scr{M}}, and are all supposed to be dynamical --- an assumption which enforces invariance under diffeomorphisms.  This framework is wide enough to cover all the metric theories of gravity.  We adopt the sign conventions of Wald's textbook~\cite{Wald:1984rg}, and use units in which the value of the speed of light in vacuum is equal to \eq{1}.

%===========================
\section{Gravitational weak equivalence principle}
\label{S:2}
%===========================

This section is devoted to reviewing the various formulations of the GWEP.  We begin by proposing a heuristic, yet physically expressive, statement of this principle, and then discuss its relation with the equality between inertial and gravitational mass.  Finally, we comment on a recent proposal for a condition that should be fulfilled by a theory in which the GWEP/SEP holds true.

%----------------------------------------
\subsection{Intuitive formulation}
\label{Ss:GWEP}
%----------------------------------------

The simplest, and most na{\"i}ve, formulation of the GWEP can be stated as follows~\cite{will, will2006lr}: {\em Freely-falling test bodies behave independently of their properties}.

Of course, this statement includes the WEP, which corresponds to the limit of negligible self-gravity.  Note that, although we allow for a sizeable self-gravity, we retain the condition that we are dealing with a test body.  The reason is that, if one considers also bodies whose back-reaction on the environment is non-negligible, it becomes impossible to compare their behaviour and then establish whether it is, or it is not, the same.  In other words, the very notion of a \emph{universal} behaviour requires that the environment remain unchanged when different bodies are considered.  By definition, this condition is accomplished provided that one works only with test bodies.

The test-body limit, however, seems hard to obtain if one takes self-gravity into account.  Consider, for example, the free fall of a micro black hole in the gravitational field of the Earth.  Of course, although the presence of the black hole does not affect the behaviour of the Earth (and, thus, the gravitational field on large scales), it does affect quite strongly the gravitational field in its surroundings.  Therefore, establishing whether the behaviour of the black hole is the same as the one of a different body (e.g., a pebble) looks problematic, because one is comparing two situations that might differ greatly in the region where motion takes place.  In other words, self-gravitating bodies can never be considered test bodies on small scales.  The way out of this deadlock is to pick an aptly chosen region surrounding the body, which extends as far as the influence of the latter becomes negligible (negligible with respect to the required precision, of course).  Then, one ignores all the details within such a region, so that the body is represented just by a world tube in a spacetime that is not affected by the body itself.

We could then try to reformulate the GWEP as the property that different bodies are represented, if they have the same initial conditions, by the same world tubes.  Unfortunately, this is still a vague statement, and it is not difficult to see that it is very hard, if not impossible, to reformulate it in a more precise way.  All difficulties disappear, however, if one focusses on very small bodies, whose world tubes can be approximated well enough by worldlines.  For all these reasons, we shall adopt the following intuitive formulation:
\begin{GWEP} 
The world lines of small freely falling test bodies do not depend on the peculiar physical properties of the bodies themselves.
\end{GWEP}
Basically, this is just an extension of the WEP, where the restriction that the particle self-gravity be negligible has been removed.  In particular, among the ``physical properties'' mentioned in the formulation of the GWEP, there can also be a parameter measuring the ratio between the gravitational self-energy and the rest energy of the body (sometimes referred to as the compactness).  Since, in a metric theory of gravity, the world-lines of non-self-gravitating test particles are the geodesics of the background geometry, the GWEP requires that this is also the case for small bodies whose self-gravity cannot be neglected.

Contrary to a widespread belief, the GWEP does not hold strictly, even in general relativity~\cite{Poisson:2011nh, Poisson:2004gg}.  Indeed, particles equipped with spin or multipole moments do not follow geodesics, because their equation of motion contains terms expressing the coupling of these quantities with the spacetime curvature~\cite{Papapetrou}.  Therefore, although these particles do not feel forces of nongravitational origin, their behaviour is not universal.  This happens even in the limit in which self-gravity can be neglected, so the violation of universality occurs already at the level of the WEP.  Moreover, taking self-gravity into account, there is a further, usually minuscule, correction that makes a particle world-line non-geodesic --- the so-called ``tail term''~\cite{Mino, QuinnWald} --- which describes the force on the body due to gravitational waves produced by the motion of the body itself at an earlier time, and then backscattered off the background curvature.\footnote{Such a term exists if one describes the particle motion with respect to the unaffected background.  However, if one adds, to the background metric, a perturbation corresponding to the ``tail'', it turns out that the particle follows a geodesic of this new metric.  Nevertheless, this does not corroborate the GWEP, because the perturbation --- and thus, the new metric --- depends on the mass of the body. Such dependance once again violates the required universality.}

Concerning these violations, we can see that those of the WEP are, after all, rather harmless, because they all follow from the internal structure of the bodies considered (a feature which survives, even in the point particle limit, in the form of quantities like the particle spin and multipole moments).  One may still argue that the WEP would hold exactly for structureless particles, although this last notion is somewhat remote from the experimental realm.  The situation is slightly less clear in the case of the tail term, for its origin is precisely the particle self-gravity, that is, the property which is at stake when it comes to test the GWEP.  It should be noted, however, that upon taking that term into account, the task of comparing the behaviour of different particles becomes hopeless, because any such comparison would demand the knowledge of the entire history of each body.  Trying to establish whether two systems behave in the same way would then require control over their entire past, which is of course impossible.  For this reason, and considering the practical irrelevance of the ``tail force'' in view of its extremely small magnitude, we shall ignore such contributions in the following analysis.  Basically, this corresponds to a restricted version of the GWEP, where one's attention is focussed on bodies whose self-radiative effects are negligible.

%----------------------------------------
\subsection{Relation with other equivalence principles}
\label{Ss:NEP}
%----------------------------------------

As seen, the GWEP nicely fits in a heterogeneous group of equivalence principles.  It is, then, worth pointing out some of the relations and logical implications among the various group members~\cite{ess1}.

As an essential request, all viable theories of gravity ought to reduce, in some limit, to Newton's theory.\footnote{Usually, this happens whenever weak gravitational fields and slow motions are considered. Nonetheless, other admissible proposals exist; for instance, the T\emph{\!e}V\emph{\!e}S theory can lead to the modified Newtonian dynamics (MOND) model; see Ref.~\cite{TeVeS} for a review.  In this latter case, however, the modifications emerge only for very small accelerations, so Newtonian dynamics still apply in a wide regime.}  At the same time, the basic principles of dynamics do not say anything about possible relations between the gravitational mass \eq{m_{\text{g}}}, and the inertial mass \eq{m_{\text{i}}}. Actually, these quantities express, in principle, two very different properties of a body.  Experiments, however, show that they are proportional to each other through a universal coefficient~\cite{will, will2006lr}, and can therefore be set equal by a suitable choice of units.  It is then fair to introduce the following additional equivalence principle, named after Newton:
\begin{NEP}[NEP] In the Newtonian limit, the inertial and gravitational masses of a particle are equal. \end{NEP}
This formulation makes it possible to test NEP also for theories other than Newton's. What really matters, in fact, is the notion of Newtonian limit, for the identification of an inertial and a gravitational mass is in general unambiguous only in those conditions.  Moreover, one can easily see that the WEP implies NEP, while the converse is not necessarily true.\footnote{The difference between the WEP and NEP is subtle, whence the frequent identification of the two statements. In general, however, the equivalence is not legitimate.  Only as long as the two masses enter in the equations of motion through their ratio \eq{m_{\text{i}} / m_{\text{g}}} alone --- as in Newton's theory --- does NEP also imply the universality of free fall, and one has WEP \eq{\Leftrightarrow} NEP.}

More to the point, the GWEP implies the WEP (and thus also NEP for all those bodies whose self-gravity is negligible).  Furthermore, the GWEP also implies NEP for all self-gravitating bodies. These last implications strongly suggest to further distinguish NEP from a \emph{gravitational NEP} (GNEP)~\cite{ess1}. It follows, then, that every time the GNEP fails, the GWEP is violated as well.  Given a theory of gravity in which the WEP holds, it is then possible to check whether NEP holds as well for a body with non-negligible self-gravity. The state-of-the-art technique to do so is the parametrized post-Newtonian (PPN) formalism~\cite{will, will2006lr}. Among its main conclusions, we underline that, although in general relativity \eq{m_{\text{i}}} always equals \eq{m_{\text{g}}}, in the Brans--Dicke theory the two masses coincide only when self-gravity can be neglected~\cite{nordtvedt}. 

That EEP implies the WEP is another straightforward conclusion. Since EEP guarantees that a freely-falling particle in a gravitational field behaves, locally, in a way indistinguishable from that of a free particle in the absence of gravity; and since the behaviour of free particles in the absence of gravity is universal, it follows that freely-falling particles also behave in a universal way.  Similarly, one can see that the SEP implies the GWEP.  The implication WEP \eq{\Rightarrow} EEP, usually referred to as Schiff's conjecture~\cite{will}, is still debated.\footnote{Interestingly enough, if Schiff's conjecture proved to be true, a connection between the GWEP and EEP could be established, stemming from the implication GWEP \eq{\Rightarrow} WEP.}

Finally, since the SEP is just EEP extended to ``test gravitational phenomena'' (of which the free fall of a test body with self-gravity is only a particular case), then it makes sense to ask whether, adding the GWEP to EEP, the full SEP could be recovered. To our knowledge, this ``gravitational extension'' of Schiff's conjecture has never been advanced before but, if correct, it would single out the GWEP as the key element of the SEP.

%-------------------------------------
\subsection{A recent approach to the SEP}
\label{Ss:gerard}
%-------------------------------------

Any principle is effective as a selection rule only as long as it can be operatively used to characterise different theories.  The GWEP and the SEP provided so far still need significant improvement in this respect.

An interesting proposal attempting to better pin down the formal content of the SEP has been advanced in Refs.~\cite{Gerard:2006ia, Gerard:2008nc}.  It builds upon the existence of an analogy between general relativity and non-Abelian gauge theories \emph{\`a la} Yang--Mills. One can then trade the usual dynamics of the metric for that of the connection (which plays the role of the gauge potential, while the curvature represents the field strength).  The ``true'' field equations thus become
\eqd{\nabla_{\! d} ?R_abc^d? = \lambda\, j_{abc} \;.
\label{gauge}}
These are second-order differential equations for the connection coefficients, sourced by ``currents'' \eq{j_{abc}} (\eq{\lambda} is a coupling constant).  Consequently, the Riemann curvature tensor is subject to the necessary condition that
\eqd{\nabla_{\! d} ?R_abc^d? = 0 \quad \text{in vacuum} \,. \label{eq:gerard}}
Since the left-hand side can be cast, in view of the Bianchi identities, in the equivalent form
\eqd{\nabla_{\! d} {R_{abc}}^d = \nabla_{\! b} R_{ac} - \nabla_{\! a} R_{bc} \;,\label{eq:rotR}}
condition~\eqref{eq:gerard} is satisfied by general relativity.  In Refs.~\cite{Gerard:2006ia, Gerard:2008nc}, it is regarded as a statement about the nonlinear character of the gravitational interaction --- ``gravitons gravitate as gluons glue'', in the words of Ref.~\cite{Gerard:2006ia}.  Condition~\eqref{eq:gerard} is also claimed to imply the SEP because, imposing it as a set of constraints on an asymptotically flat, spherically symmetric metric representing a weak stationary field, one finds exactly the conditions on the PPN parameters that guarantee the validity of the NEP for self-gravitating bodies~\cite{nordtvedt}.  

This proposal has for sure the merit of trying to turn the SEP into a  formally precise statement.  It must be stressed, however, that the PPN constraints emerging from condition~\eqref{eq:gerard} only provide a test of the GNEP for bodies with weak self-gravity~\cite{nordtvedt}.  Unless this implies the SEP, it is not obvious that Eq.~\eqref{eq:gerard} can be taken as a sufficient condition for the SEP.

Furthermore, while it is proven that Eq.~\eqref{eq:gerard} fails for the Brans--Dicke theory (as expected from a test of the SEP), an extension to other theories of gravity does not appear at all straightforward, for no general prescription is offered to build the equivalent of Eq.~\eqref{eq:gerard} in the case of, e.g., multi-scalar-tensor theories, higher-curvature theories and so forth.

For all these reasons, we shall try now to make a further step ahead, and propose a precise, independent formulation of the GWEP which can also be operatively useful.

%===========================
\section{A criterion for the GWEP}
\label{S:5}
%===========================

The formulation of the GWEP presented in Sec.~\ref{Ss:GWEP} offers a natural procedure for probing the validity of the principle for a metric theory of gravity.  Given any test particle, treated as non-self-gravitating, we consider its world line on an assigned arbitrary background, and then try to prove that, when self-gravity is ``turned on'', its new world-line coincides with the previous one. 

Such criterion, it is worth stressing again, is intrinsically perturbative in nature, for one ends up checking that a small body --- treated alternatively as non-self-gravitating and self-gravitating --- will basically follow the same path on a given background spacetime.  We are implicitly conjecturing, then, that the perturbation induced by ``switching on'' the self-gravitation of the body can be suitably neglected outside an appropriate world tube around the test body, or, stated otherwise, that the self-gravitating and the non-self-gravitating body live on the \emph{same} background.  This point is crucial, as it allows to sidestep all the complications related to possible effects of the body on the surrounding geometry, and hence it relieves completely the ambiguities associated with the expression ``behaving in the same way''.

The computation of the world-tube for a self-gravitating body is, in general, quite a difficult task. Nonetheless, as just stressed, all we need is an acid test of the fact that the world tube of a self-gravitating body contains a geodesic of the original background spacetime, rather than a detailed description of it.  Such a test is viable, and is facilitated by a well-known theorem.

By appropriately deploying the content of such proposition, the perturbative approach intrinsic to the problem, and finally the variational formulation of a metric theory of gravity, it becomes possible to give the heuristic criterion a fully formal status, allowing for a straightforward check of its validity.

%----------------------------------------
\subsection{Motion of a small body}
\label{Ss:gje}
%----------------------------------------

In 1975, Geroch and Jang proved the following statement, refined some years later by Ehlers and Geroch ~\cite{Geroch:1975uq, Ehlers:2003tv}:\footnote{The formulation of the theorem presented here is excerpted from Malament \cite{Malament:2009ee}, where interesting remarks can be found about the niceties of the proposition and its hypotheses.}
\begin{GJ}
Let \eq{\Scr{C}} be a smooth curve in a spacetime \eq{\ton{\Scr{M}, \matg_{ab}}}. Suppose that, given any open subset \eq{\Scr{U}} of \eq{\Scr{M}} containing \eq{\Scr{C}}, there exists a smooth symmetric field \eq{\Theta_{ab}} on \eq{\Scr{M}} such that: (a) \eq{\Theta_{ab}} satisfies the strengthened dominant energy condition;\footnote{That is, for all points \eq{P \in \Scr{M}}, and for all unit time-like vectors \eq{\xi^a} at \eq{P}, it is \eq{\Theta_{ab}\, \xi^a \xi^b \ge 0}, and, if \eq{\Theta_{ab} \neq 0}, then \eq{{\Theta^a}_b\, \xi^b} is time-like.} (b) \eq{\Theta_{ab} \neq 0} at some point in \eq{\Scr{U}}; (c) \eq{\Theta_{ab} = 0} outside of \eq{\Scr{U}}; (d) \eq{\nabla^b \Theta_{ab} = 0}.  Then, \eq{\Scr{C}} is a timelike geodesic on \eq{\ton{\Scr{M}, \matg_{ab}}}.
\end{GJ}
This seems to be exactly what we need to implement the universality of free fall.  Indeed, the stress-energy-momentum tensor \eq{T_{ab}} of a localised body satisfies conditions (a)--(d) above.\footnote{Of course, condition d) holds only if no forces other than gravity act on the body, which is precisely the case in free fall experiments.}  Hence, its world tube always contains a timelike geodesic, to which it can be approximated when the spatial extension of the body can be neglected.

Yet, this conclusion is not entirely satisfactory, because it is easy to see that conditions (a)--(d) hold independently of whether the body is, or is not, self-gravitating.  The world-tube of a very small body can then always be approximated by a geodesic, and the GWEP seems to be satisfied trivially.  Moreover, it appears that there is no substantial difference between the WEP and the GWEP, because they both hold under exactly the same conditions. Such a conclusion looks suspicious, to say the least.

A little thinking shows, however, that, although it is true that the Geroch--Jang theorem implies geodesic motion for both a non-self-gravitating and a self-gravitating small body, such geodesics belong to \emph{different} spacetimes.  In the former case, the world tube approximates a geodesic in a background spacetime that does not satisfy the gravitational field equations with the body acting as a source.  In the latter case, on the other hand, the world tube approximates a geodesic of the spacetime where the body is included among the sources of gravity --- a spacetime varying from one body to another.  Therefore, while the Geroch--Jang theorem suffices to prove the validity of the WEP, it does not imply that the GWEP also holds.  As we pointed out in Sec.~\ref{Ss:GWEP}, for this purpose one ought to show that small self-gravitating bodies all have the same set of world lines in a background spacetime which is not affected by their presence.  Since the stress-energy-momentum tensor \eq{T_{ab}} satisfies condition (d) when \eq{\nabla} is the covariant derivative with respect to the ``full'' spacetime, i.e., the one that solves the gravitational field equations considering the body itself as a source, but not with respect to the background, one cannot conclude that the universality of free fall automatically extends to self-gravitating bodies as well.

The GWEP would still hold, nevertheless, if one could find \emph{another} symmetric tensor \eq{\Theta_{ab}}, satisfying conditions (a)--(d) in the background.  In particular, we stress that condition (d) should take the form
\eqd{\bar{\nabla}^b \Theta_{ab}=0\;, \label{quast}}
where the covariant derivative \eq{\bar{\nabla}}, and the metric \eq{\bar{\matg}_{ab}} used to raise the index, now both refer to the background geometry.  Of course, for a self-gravitating body, such a \eq{\Theta_{ab}} cannot coincide with the stress-energy tensor \eq{T_{ab}}.  It is plausible that, if such a \eq{\Theta_{ab}} exists at all, it should include also a contribution associated with self-gravitation.

As we shall see below, a perturbative approach naturally leads to a reasonable candidate for \eq{\Theta_{ab}} --- even better, one satisfying condition~\eqref{quast} in the case of general relativity, but not in the case of other theories that are known to violate the GWEP.  In fact, whether the \eq{\Theta_{ab}} emerging from the perturbative expansion does, or does not, satisfy condition~\eqref{quast}, can be traced back to a specific property of the variational formulation of the  theory of gravity under consideration.  The tough problem of checking the validity of the GWEP is thus reduced to a much simpler one, namely, the mere inspection of the action functional.

Remarkably, the Geroch--Jang theorem does not depend on the detailed field equations for the metric, nor on the possible presence of additional gravitational degrees of freedom (e.g., the scalar field in the Brans--Dicke theory), nor on the number of dimensions of  spacetime.  Therefore, it is a promising starting point to test whether a generic metric theory of gravity satisfies or not the GWEP.

Before we turn to the perturbative scheme, we must still discuss two issues potentially threatening the entire construction.  First: the Geroch--Jang theorem implies that, if \eq{\Theta_{ab}} satisfying conditions (a)--(d) exists, then the world line of a small body is a geodesic, hence the universality of free fall.  Imagine, however, that we can find \eq{\Theta_{ab}} satisfying (a)--(c) but not (d). What can be said about the resulting motion of the body?

Strictly speaking, we cannot claim that relaxing hypothesis (d) implies a nongeodesic path, because conditions (a)--(d) are only sufficient for the theorem to hold.  Nevertheless, if one has good reasons to believe that \eq{\Theta_{ab}} gives an appropriate description of the total energy-momentum content of the body --- in the sense that its integral over a spacelike hypersurface represents the body four-momentum \eq{p_a} --- then one can adapt a  derivation in Ref.~\cite{QuinnWald}, pp.~3382--3383, and argue that, if \eq{\bar{\nabla}^b \Theta_{ab} \neq 0}, then an external force acts on the body and prevents its world line from being a geodesic.  For this reason, in the rest of the paper we shall assume that the GWEP holds \emph{if and only if} condition~\eqref{quast} is satisfied, in addition to (a)--(c), for a tensor \eq{\Theta_{ab}} such that the difference \eq{\Theta_{ab} - T_{ab}} represents the stress-energy-momentum associated with the gravitational field of the body under consideration. 

The second issue deals with condition (c).  If the body is spatially localised, then \eq{T_{ab}} vanishes outside a world tube.  But this is not the case for \eq{\Theta_{ab}}, which contains also a contribution from the gravitational field.  The dominant part of this contribution is the energy density of the static gravitational field associated with the body, which falls off as \eq{r^{-4}}, where \eq{r} is a suitable radial coordinate.  Since this function has not a compact support, \eq{\Theta_{ab}} fails to satisfy (c).  To fix such issue, one has to notice that, in the proof of the Geroch--Jang theorem, condition (c) is used to guarantee that the integral over a spacelike hypersurface of a particular function proportional to \eq{\Theta_{ab}} reduces to an integral performed only over the volume of the body.  Calling \eq{M} the mass of the body, and \eq{R} the spatial size of the world tube associated with it, the contribution due to the external static gravitational field turns out to be proportional to \eq{{G M^2}/R}.  For a non-compact body, this is  much smaller than the proper energy \eq{M c^2}, and even for an extreme object like a black hole it is sufficient to choose \eq{R} much larger that the Schwarzschild radius --- but still so small that the world tube can be well approximated by a world line --- to ensure that no appreciable error arises when one truncates the range of integration at \eq{R}.  Therefore, in the following, we shall assume that a weaker form of the Geroch--Jang theorem holds, one which guarantees geodesic motion even under these conditions.\footnote{Even worse, there might be radiative contributions, whose energy density falls off as \eq{r^{-2}}.  As already pointed out in Sec.~\ref{Ss:GWEP}, we shall always restrict ourselves to experimental conditions where such terms can be neglected, so that the GWEP is at least satisfied by general relativity.}

%----------------------------------------
\subsection{Self-gravitation: The perturbative approach}
\label{Ss:sg}
%----------------------------------------

We can now exhibit the actual tensor \eq{\Theta_{ab}} required to test the validity of the GWEP for a given metric theory of gravity. To do so, we first observe that every such theory contains field equations that can always be cast in the form\footnote{The very structure of our test for the GWEP makes it impossible to apply it also to a scalar theory of gravity. This leaves outside Nordstr{\"o}m's theory, which in turn is known to abide at least by the GNEP~\cite{deruelle}. The reason is that, in such a theory, the dynamical scalar degree of freedom lives on a fixed, nondynamical minkowskian background; this conflicts with our assumptions that only truly dynamical fields, viz. diffeomorphism-invariant structures, are allowed. As a consequence, one has that the equations of motion for test bodies cannot be derived from the field equations --- as it happens, e.g., in general relativity~\cite{infeldschild}, in view of its fully dynamical character --- but must be instead postulated separately~\cite{Ravndal:2004ym}. This issue, together with the impossibility of checking condition (d) in the Geroch--Jang theorem, prevents us from applying our method to a purely scalar theory of gravity.}
\eqd{E_{ab} = T_{ab} \;. \label{eq:Gen-eq} }
Here, the stress-energy-momentum tensor \eq{T_{ab}} can depend on the metric field \eq{\matg_{ab}} and any matter field, but not on other possible gravitational degrees of freedom.\footnote{In the context of scalar-tensor theories (see Sec.~\ref{ss:scalartens} below), this is tantamount to choosing the so-called ``Jordan frame''.  In this way, non-self-gravitating particles follow the geodesics of the metric, and the WEP is satisfied --- clearly, a prerequisite for the GWEP.}  On the other hand, the tensor \eq{E_{ab}} depends on all the gravitational variables and their derivatives.  In the particular case of general relativity, for instance, it is 
\eqd{E_{ab} = \frac{1}{8 \pi G} \, G_{ab} \;, \label{EGR}}
where \eq{G_{ab}} is the Einstein tensor.  If gravity is uniquely described by \eq{\matg_{ab}}, we say that the theory is \emph{purely metric}.  If this is not the case, there are other gravitational field equations in addition to~\eqref{eq:Gen-eq}, associated with the other gravitational variables.  Finally, there are the matter field equations.

Let us now suppose that a solution \eq{\matg_{ab}} of Eq.~\eqref{eq:Gen-eq} can be decomposed into the sum of two terms, namely a ``background'' \eq{\bar{\matg}_{ab}}, and a ``small perturbation'' \eq{\epsilon\, \gamma_{ab}}, where \eq{\epsilon} is a bookkeeping  parameter accounting for the correct orders of \eq{\gamma_{ab}} in the expansions, so that
\eqd{\matg_{ab} = \bar{\matg}_{ab} + \epsilon \, \gamma_{ab} \label{eq:metricexp}}
(see Appendix~\ref{App:A} for the corresponding expansion of several geometric quantities of interest). We can expand the tensors \eq{E_{ab}} and \eq{T_{ab}} accordingly, obtaining
\subeq{\gat{E_{ab} = \bar{E}_{ab} + \epsilon \, \Cal{E}_{ab} + E^{(2+)}_{ab} \,, \label{Eexp}\\
T_{ab} = \bar{T}_{ab} + \epsilon \, \Cal{T}_{ab} + T^{(2+)}_{ab} \,. \label{eq:expT}}
\label{eq:expand}}
From this point onwards, symbols with an over bar are associated with quantities constructed out of background objects only.  The script letters denote the parts which are linear in the perturbations, whereas the superscript ``\eq{(2+)}'' generically denotes all the terms of order higher or equal than two in the expansions~\cite{Mino}. We assume furthermore that the background metric satisfies the zeroth-order field equations
\eqd{\bar{E}_{ab} = \bar{T}_{ab} \;. \label{eq:zerothback}}

As it is normally done when testing the GWEP, we set  \eq{\bar{T}_{ab} = 0}; that is, we ask the background solution to be a vacuum one; the crucial importance of such assumption will be fully explained at the end of Sec.~\ref{Ss:var}.  We can thus be sure that \eq{T_{ab}} is made entirely of the small contribution due to the stress-energy-momentum tensor \eq{T^{(p)}_{ab}} of the ``particle'' (i.e., the self-gravitating test body).  Then, we can recast the field equations~\eqref{eq:Gen-eq} in the equivalent form
\eqd{\epsilon \, \Cal{E}_{ab} = T^{(p)}_{ab} - E^{(2+)}_{a b}.
\label{eq:estorto}}
If now one can prove that
\eqd{\bar{\nabla}^b \Cal{E}_{ab} = 0 \label{eq:cons} \,,}
then the term \eq{E^{(2+)}_{a b}} can be considered as a contribution due to  the gravitational self-interaction of the body, and can be included into a new symmetric tensor
\eqd{\Theta_{ab} =  T^{(p)}_{ab} - E^{(2+)}_{a b} \;,
\label{eq:Theta}}
which is conserved with respect to the background in view of Eq.~\eqref{eq:cons}. This is all we need to invoke the Geroch--Jang theorem,\footnote{Strictly speaking, one should also check  that \eq{\Theta_{ab}} satisfies the other hypotheses in the theorem.  However, \eq{T_{ab}} satisfies them by assumption.  Moreover, as discussed in Sec.~\ref{Ss:gje}, we suppose that the perturbation is sufficiently weak so that \eq{E^{(2+)}_{a b}} does not alter condition (a), and that condition (c) can be relaxed without altering the conclusion.} and  hence deduce that the self-gravitating body is moving within a world tube which can be approximated with arbitrary precision by a geodesic of the background.

For all these reasons, we consider Eq.~\eqref{eq:cons} to be a \emph{necessary and sufficient} condition for the GWEP to hold.

%------------------------------------------
\subsection{Variational formulation of the GWEP-test}
\label{Ss:var}
%------------------------------------------

The basic ingredients of our method --- namely, equations~\eqref{eq:Gen-eq}, \eqref{Eexp}, and~\eqref{eq:cons} --- might suggest at this stage that the actual check of the validity of the GWEP becomes only a matter of case-by-case calculations. For each metric theory of gravity,  the different form of the tensors \eq{E_{ab}} and \eq{\Cal{E}_{ab}} may in principle demand a separate analysis. However, we shall see immediately that such a huge labour is not necessary, and can be effectively traded for a much more elegant and quicker inspection of the action of a given theory.

To this end, we notice first that, in a variational formulation of a theory of gravity, the tensors \eq{E_{ab}} and \eq{T_{ab}} in Eq.~\eqref{eq:Gen-eq} are given by the functional derivatives, with respect to the inverse metric \eq{\matg^{ab}}, of the gravitational action \eq{A_{\text{G}}}, and of the matter action \eq{A_{\text{M}}}, respectively:
\eqd{E_{ab} = \frac{2}{\sqrt{-\matg}} \, \frac{\delta A_{\text{G}}}{\delta \matg^{ab}}\;; \quad  \quad T_{ab} = - \frac{2}{\sqrt{-\matg}} \, \frac{\delta A_{\text{M}}}{\delta \matg^{ab}} \;.}
Since the framework we are building is completely general, \eq{A_{\text{G}}} is in principle a functional of the metric \eq{\matg_{ab}}, and of any other gravitational degree of freedom \eq{\Phi_I} (\eq{I} represents a convenient set of indices).  At the same time, the matter action \eq{A_{\text{M}}} is a functional of \eq{\matg_{ab}} and of the matter fields, but \emph{not} of the \eq{\Phi_I}, see the footnote commenting Eq.~\eqref{eq:Gen-eq}.

The field equations~\eqref{eq:Gen-eq} then follow as soon as one requires that the total action \eq{A = A_{\text{G}} + A_{\text{M}}} has a vanishing functional derivative with respect to \eq{\matg^{ab}}.   The matter field equations emerge from \eq{A_{\text{M}}} alone when varied with respect to the matter fields.  Similarly, additional gravitational field equations are derived varying \eq{A_{\text{G}}} with respect to the \eq{\Phi_I}. It is worth remarking here that the action \eq{A} needs be equipped with all the apt boundary terms, necessary to derive the field equations in a consistent manner --- that is, to guarantee that the functional derivatives of \eq{A}, with respect to the various fields, exist.  In the case of general relativity, for instance, one has to add the Gibbons--Hawking--York term~\cite{Gibbons:1976ue, York:1972sj}.

We can now decompose all the dynamical variables into a background part and a perturbation, the latter being associated with the parameter \eq{\epsilon}, as in Sec.~\ref{Ss:sg}.  The gravitational action can then be rewritten as 
\eqd{A_{\text{G}} = \bar{A}_{\text{G}} + \epsilon \, \Cal{A}_{\text{G}} + A_{\text{G}}^{(2+)}\;,
\label{AAAA}}
where the zeroth-order term \eq{\bar{A}_G} is a functional of the background fields only, whereas \eq{\Cal{A}_G} is a functional of the background and of the first-order perturbations.  For convenience, we shall consider, as independent variables, the tensors \eq{\bar{\matg}^{ab}} and \eq{\gamma^{ab}}.  This choice is entirely equivalent to the one in which the independent variables are \eq{\bar{\matg}_{ab}} and \eq{\gamma_{ab}}, but simplifies the treatment.

Varying Eq.~\eqref{AAAA} with respect to the background metric, and noticing that \eq{\gamma^{ab}} is independent of \eq{\bar{\matg}^{ab}}, one finds
\eqd{\frac{\delta \bar{A}_{\text{G}}}{\delta \bar{\matg}^{ab}} + \epsilon \, \frac{\delta \Cal{A}_{\text{G}}}{\delta \bar{\matg}^{ab}} + \frac{\delta A^{(2+)}_{\text{G}}}{\delta \bar{\matg}^{ab}} = \frac{\delta A_{\text{G}}}{\delta \bar{\matg}^{ab}} = \frac{\delta A_{\text{G}}}{\delta \matg^{cd}} \, \frac{\partial \matg^{cd}}{\partial\bar{\matg}^{ab}} = \frac{\delta A_{\text{G}}}{\delta \matg^{ab}} \;.\label{bibi}} 
On the other hand, to the first order in \eq{\epsilon}, it is
\eqd{\frac{\delta A_{\text{G}}}{\delta \matg^{ab}} = \frac{\sqrt{-\matg}}{2} \, E_{ab} = \frac{\sqrt{-\bar{\matg}}}{2} \, \bar{E}_{ab} + \epsilon \, \frac{\sqrt{-\bar{\matg}}}{2} \ton{\frac{\gamma}{2} \, \bar{E}_{ab} + \Cal{E}_{ab}} , \label{bibo}}
where we have used Eqs.~\eqref{Eexp} and~\eqref{ggbar}, and set \eq{\gamma := \bar{\matg}_{ab} \gamma^{ab}}.  Comparing Eqs.~\eqref{bibi} and~\eqref{bibo} we find
\subeq{\gat{\bar{E}_{ab} = \frac{2}{\sqrt{- \bar{\matg}}} \, \frac{\delta \bar{A}_{\text{G}}}{\delta \bar{\matg}^{ab}} \;;\\
\Cal{E}_{ab} = \frac{2}{\sqrt{-\bar{\matg}}} \, \frac{\delta \Cal{A}_{\text{G}}}{\delta \bar{\matg}^{ab}} - \frac{\gamma}{2}\,\bar{E}_{ab}\;.
\label{chicchirichi}}}
For the sake of definiteness, and without loss of generality, we suppose that all the non-metrical gravitational degrees of freedom reduce for the moment to a single scalar field \eq{\phi} (the treatment can be straightforwardly  extended to other geometric objects), and perform an analogous expansion
\eqd{\phi = \bar{\phi} + \epsilon\,\chi + \phi^{(2+)} \;.
\label{eq:phiexp}}
Then, \eq{\bar{A}_{\text{G}} \qua{\bar{\matg}^{ab} , \bar{\phi}} = A_{\text{G}} \qua{\bar{\matg}^{ab} , \bar{\phi}}}, and since the matter action does not depend on \eq{\phi}, the background scalar field \eq{\bar{\phi}} must obey the equation 
\eqd{\frac{\delta \bar{A}_{\text{G}}}{\delta\bar{\phi}} = 0 \;. \label{phiback}}
Interestingly, it is also:
\subeq{\eqd{\frac{\delta \Cal{A}_{\text{G}}}{\delta \gamma^{ab}} = - \frac{\sqrt{-\bar{\matg}}}{2} \, \bar{E}_{ab}\;; \label{strana}}
and
\eqd{\frac{\delta \Cal{A}_{\text{G}}}{\delta \chi} = \frac{\delta \bar{A}_{\text{G}}}{\delta \bar{\phi}} \;. \label{stranaphi}}}
These relations appear obvious, as soon as one notices that \eq{\epsilon \, \Cal{A}_{\text{G}}} is just the first-order variation of \eq{A_{\text{G}}} corresponding to the variations \eq{\delta\matg^{ab} = - \epsilon \, \gamma^{ab}} and \eq{\delta \phi = \epsilon \, \chi} of the fields.  It is instructive, however, to see a formal proof.  An explicit calculation of \eq{\Cal{A}_{\text{G}}} gives
\begin{widetext}
\speq{\Cal{A}_{\text{G}} \qua{\bar{\matg}^{ab} , \gamma^{ab} , \bar{\phi} , \chi} &=\ton{\diff{A_{\text{G}} \qua{\matg^{ab} , \phi}}{\epsilon}}_{\!\! \epsilon = 0} = \int_{U} \de^m x \, \qua{\ton{\frac{\delta A_{\text{G}}}{\delta \matg^{ab}}}_{\!\! \epsilon = 0} \ton{\diff{\matg^{ab}}{\epsilon}}_{\!\! \epsilon = 0} + \ton{\frac{\delta A_{\text{G}}}{\delta \phi}}_{\!\! \epsilon = 0} \ton{\diff{\phi}{\epsilon}}_{\!\! \epsilon = 0}\,} \\
& =\int_{U} \de^m x\, \ton{- \frac{\delta \bar{A}_{\text{G}}}{\delta \bar{\matg}^{ab}}\,\gamma^{ab} + \frac{\delta \bar{A}_{\text{G}}}{\delta \bar{\phi}} \, \chi} = \int_{\Scr{U}} \de \bar{\Omega} \, \ton{- \frac{1}{2} \, \bar{E}_{ab} \, \gamma^{ab} + \frac{1}{\sqrt{-\bar{\matg}}} \, \frac{\delta \bar{A}_{\text{G}}}{\delta \bar{\phi}} \, \chi} \;, \label{ponzipo}}
\end{widetext}
where \eq{\Scr{U}} and \eq{U} denote, respectively, the domain of integration and its coordinate representation in \eq{\bol{R}^m} (for simplicity, we assume that \eq{\Scr{U}} can be covered by a single chart), and \eq{\de \bar{\Omega} = \sqrt{-\bar{\matg}} \; \de^m x} is the volume element with respect to the background metric. Equations~\eqref{strana} and~\eqref{stranaphi} then follow immediately.

It is important to realise that this expression for \eq{\Cal{A}_{\text{G}}} is correct provided that the functional derivatives of \eq{A_{\text{G}}} with respect to the gravitational fields \eq{\matg^{ab}} and \eq{\phi} exist.   As is well-known, this is the case provided that suitable conditions on the fields themselves and their derivatives are imposed on the boundary \eq{\partial \Scr{U}}.  Of course, this basically amounts to requiring that the variational principle be well-formulated for the theory under scrutiny.  

Consider now a diffeomorphism, generated by the vector field \eq{\xi^a}.  Order by order in \eq{\epsilon}, diffeomorphism invariance of the gravitational theory implies \eq{\delta\bar{A}_{\text{G}} = 0}, \eq{\delta \Cal{A}_{\text{G}} = 0}, and so forth.  In particular, the second of these conditions becomes, with standard manipulations, 
\begin{widetext}
\speq{0 &= \int_{U} \de^m x \, \ton{\frac{\delta \Cal{A}_{\text{G}}}{\delta \bar{\matg}^{ab}} \, \delta \bar{\matg}^{ab} + \frac{\delta \Cal{A}_{\text{G}}}{\delta \gamma^{ab}} \,\delta \gamma^{ab} + \frac{\delta \Cal{A}_{\text{G}}}{\delta \bar{\phi}} \, \delta \bar{\phi} + \frac{\delta \Cal{A}_{\text{G}}}{\delta \chi} \, \delta \chi} \\
&=\int_{\Scr{U}} \de \bar{\Omega}\, \ton{-\frac{2}{\sqrt{- \bar{\matg}}} \, \frac{\delta  \Cal{A}_{\text{G}}}{\delta \bar{\matg}^{ab}} \, \bar{\nabla}^b \xi^a + \frac{1}{\sqrt{- \bar{\matg}}} \, \frac{\delta \Cal{A}_{\text{G}}}{\delta \gamma^{ab}} \, \delta \gamma^{ab} + \frac{1}{\sqrt{-\bar{\matg}}} \, \frac{\delta \Cal{A}_{\text{G}}}{\delta \bar{\phi}} \, \xi^a \, \bar{\nabla}_{\! a} \bar{\phi}} \;,}
where \eq{\delta \gamma^{ab} = \xi^c \bar{\nabla}_{\! c} \gamma^{ab} - \gamma^{cb} \bar{\nabla}_{\! c} \, \xi^a - \gamma^{ac} \bar{\nabla}_{\! c} \, \xi^b}, and we have dropped a term using  Eq.~\eqref{stranaphi} together with the field equation~\eqref{phiback} for the background scalar field \eq{\bar{\phi}}.  Integrating by parts, and eliminating a boundary term choosing the arbitrary vector field \eq{\xi^a} such that it vanishes on \eq{\partial \Scr{U}}, we find
\speq{0 &= \int_\Scr{U} \de \bar{\Omega} \qua{\ton{\bar{\nabla}^b \ton{\frac{2}{\sqrt{- \bar{\matg}}} \, \frac{\delta \Cal{A}_{\text{G}}}{\delta \bar{\matg}^{ab}}} + \frac{1}{\sqrt{- \bar{\matg}}} \, \frac{\delta \Cal{A}_{\text{G}}}{\delta \bar{\phi}} \, \bar{\nabla}_{\! a}\bar{\phi} } \xi^a + \frac{1}{\sqrt{- \bar{\matg}}} \, \frac{\delta \Cal{A}_{\text{G}}}{\delta \gamma^{ab}} \, \delta \gamma^{ab}} \\
&= \int_{\Scr{U}} \de \bar{\Omega}\, \ton{\bar{\nabla}^b \Cal{E}_{ab} + \frac{1}{\sqrt{- \bar{\matg}}} \, \frac{\delta \Cal{A}_{\text{G}}}{\delta \bar{\phi}} \, \bar{\nabla}_{\! a}\bar{\phi} - \frac{1}{2} \, \bar{E}_{bc} \bar{\nabla}_{\! a} \gamma^{bc} - \bar{\nabla}_{\! c} (\bar{E}_{ab} \, \gamma^{cb}) + \frac{1}{2}\,\bar{\nabla}^b(\gamma\,\bar{E}_{ab})} \xi^a\;,}
where we have used Eqs.~\eqref{chicchirichi} and~\eqref{strana}.  By the arbitrariness of \eq{\xi^a} we find, finally, 
\eqd{\bar{\nabla}^b \Cal{E}_{ab} = -\frac{1}{\sqrt{- \bar{\matg}}} \, \frac{\delta \Cal{A}_{\text{G}}}{\delta \bar{\phi}} \, \bar{\nabla}_{\! a} \bar{\phi} + \frac{1}{2} \, \bar{E}_{bc} \bar{\nabla}_{\! a} \gamma^{bc} +  \bar{\nabla}_{\! b} (\bar{E}_{ac} \gamma^{bc})-\frac{1}{2}\,\bar{E}_{ab}\bar{\nabla}^b\gamma\;.
\label{popopoff}}
\end{widetext}
In the last term above, we have used the property \eq{\bar{\nabla}^b\bar{E}_{ab}=0}, which follows by diffeomorphism invariance of \eq{\bar{A}_{\text{G}}} together with the field equation~\eqref{phiback} for \eq{\bar{\phi}}.

Since \eq{\bar{\phi}} and \eq{\gamma_{ab}} are independent, equation~\eqref{popopoff} implies that, in order for Eq.~\eqref{eq:cons} to hold, the following two conditions must be satisfied simultaneously:\footnote{These conditions are unavoidable.  In particular, one might try to escape condition~\eqref{ii)} making the last three terms in Eq.~\eqref{popopoff} vanish, exploiting the gauge freedom \eq{\gamma_{ab} \to  \gamma_{ab} + \bar{\nabla}_{\! a} \zeta_{b}+ \bar{\nabla}_{\! b} \zeta_{a}}, where \eq{\zeta_a} is arbitrary.  Of course, this will not succeed, because the independent terms to be killed are  more than the four arbitrary functions provided by \eq{\zeta_a}.  Moreover, such a transformation would also affect \eq{\Cal{E}_{ab}}, thus making other terms appear on the left-hand side of Eq.~\eqref{popopoff}.}
\subeq{\gat{\frac{\delta \Cal{A}_{\text{G}}}{ \delta \bar{\phi}} = 0\;; \label{i)} \\
\bar{E}_{ab} = 0 \;. \label{ii)}}}
These are rather different in character, so we comment on them separately.

Condition~\eqref{i)} is ``structural'', i.e., it concerns the type of theory considered.  One can realise that, even assuming the validity of the field equations, the quantity \eq{{\delta\Cal{A}_{\text{G}}}/{\delta\bar{\phi}}} vanishes only if the gravitational action does not depend on the variable \eq{\phi}. Dropping the initial restriction of a single additional scalar field, this is tantamount to requiring that one is dealing with a purely metric theory of gravity.  Of course, a constant background scalar field \eq{\bar{\phi}} would still make the first term on the right-hand side of Eq.~\eqref{popopoff} vanish.  This circumstance, however, can occur only for extremely special background metrics, hence it is irrelevant for the general validity of the GWEP within a given theory.

On the other hand, condition~\eqref{ii)} is ``environmental''.  Using the field equations~\eqref{eq:zerothback}, it implies that condition~\eqref{eq:cons} can be satisfied only in an empty background (that is, for \eq{\bar{T}_{ab} = 0}).  This assumption is always made in the context of tests of the GWEP or of the SEP, and it is interesting to find that it emerges from our formalism as a necessary condition for the GWEP.  Let us then clarify its role.

Suppose that \eq{\bar{T}_{ab} \neq 0} for some otherwise unspecified background.  A self-gravitating body produces, by definition, a non-negligible gravitational field in its surroundings, and this exerts, in general, a gravitational force on the background matter.  By the action-reaction principle, there is also a corresponding force on the body itself, which makes its world line deviate from a geodesic of \eq{\bar{\matg}_{ab}}.  A Newtonian estimate for the potential energy density associated with these forces is \eq{\bar{\rho} \, \Phi}, where \eq{\Phi} is the gravitational potential due to the body, and \eq{\bar{\rho}} is the mass density of the background.  The relativistic tensorial generalisation of \eq{\bar{\rho} \, \Phi} is \eq{\bar{T}_{ac} \, \gamma^{cb}}, and considering the field equations~\eqref{eq:zerothback} for the background,  this explains the appearance of the terms in Eq.~\eqref{popopoff}.  The derivatives are, of course, due to the fact that the quantity evaluated in Eq.~\eqref{popopoff} is related to the force density, rather than to the potential energy density.

To sum up, an empty background is a necessary condition for the GWEP to hold, because in the presence of matter the test body couples to it via its own gravitational field, and this produces, in general, a force that violates the universality of free fall.  Thus, \eq{\bar{T}_{ab} = 0} is one of the basic conditions (together with the assumptions that the body is spatially localised, and that it affects the environment negligibly) that must be postulated so that the very issue of whether a theory does, or does not, satisfy the GWEP makes sense at all.

This conclusion receives additional, independent support if one notices that the presence of matter at the background level would prevent the identification \eq{T_{ab} = T^{(p)}_{ab}}, which was crucial for writing Eq.~\eqref{eq:estorto}.  This would imply that one is no longer able to confine the generalised stress-energy-momentum tensor \eq{\Theta_{ab}} given by Eq.~\eqref{eq:Theta} to the interior of a world tube, as it will depend not only on the particle, but in general also on the distribution of background matter.  This fact in turn violates one of the assumptions required for the Geroch--Jang theorem to hold, thus shaking the very foundations of our framework.

%===========================
\section{Case studies}
\label{S:6}
%===========================

We now investigate explicitly whether condition~\eqref{eq:cons} for the GWEP holds in some relevant classes of metric theories.  This analysis complements the general results obtained in Sec.~\ref{Ss:var}, culminating in Eq.~\eqref{popopoff}, and allows for a further discussion of some niceties.  We first consider two cases --- general relativity and the Brans--Dicke theory --- in which the validity of the GWEP has been already confirmed or disproved by means of other theoretical and experimental arguments.  In this sense, such theories are used here as preliminary sanity checks.

After that, we move on to three other cases where the status of the principle is less certain, or even completely unknown.  Specifically, we examine general scalar-tensor theories, then higher-curvature and higher-order \eq{f(R)} theories, and finally the higher-curvature but not higher-order Lanczos--Lovelock theories.  The flexibility and generality of condition~\eqref{eq:cons}, together with the variational formalism, will show all its power precisely in such more exotic scenarios, where other theoretical arguments turn out to be inadequate or even completely inapplicable.\footnote{The standard PPN formalism, for instance, is tailored on a four-dimensional spacetime, and becomes of no practical utility (unless  suitably extended) in higher dimensions, which is what happens precisely in the context of Lanczos--Lovelock theories of gravity.}

The simple ``yes/no'' answer one can extract from a test of the GWEP, however, is not the only relevant bit of information one can get.  By this analysis, we can also try to better understand in which precise sense the fact that a given theory satisfies or not the GWEP has to be linked with the existence of additional gravitational degrees of freedom, as it happens in the Brans--Dicke theory, or can be also attributed to the presence of higher-curvature terms in the action, as in \eq{f(R)} and the Lanczos--Lovelock theories.

%--------------------------------------
\subsection{General relativity}
\label{ss:GR}
%--------------------------------------

Being the test-bench upon which the SEP (and, thus, the GWEP) has been tailored, it is evident that Einstein's tensor theory of gravity in four spacetime dimensions abides by such principle --- of course, with all the caveats discussed in Secs.~\ref{Ss:GWEP} and~\ref{Ss:gje}. However, given the features of our protocol, the restriction to four-dimensional manifolds can be set aside, and we can test the validity of the GWEP on any \eq{m}-dimensional version of general relativity (with \eq{m \ge 4}). Hence, in the present section we work directly in the most general case.

The Einstein--Hilbert action (with the addition of the Gibbons--Hawking--York boundary term~\cite{York:1972sj, Gibbons:1976ue, Dyer:2008hb}) underlying the gravitational sector of the theory is
\eqd{A_{\text{GR}} = \frac{1}{16\pi G} \ton{\int_{\Scr{U}} \!\! \de \Omega \, \ton{R  - 2 \Lambda} + 2 \oint_{\pd \Scr{U}} \!\!\! \de \Sigma \, K} . \label{eq:einhilact}}
Here, \eq{R} is the scalar curvature, \eq{\Lambda} is a cosmological constant\footnote{The case of a \eq{\Lambda} term has not been investigated thoroughly using the PPN formalism, mostly because of its extremely tiny observational value (of order \eq{10^{-122}} in Planck units), which is usually considered small enough to be neglected in the context of solar system experiments.  Hence, it remains unclear whether the addition of such a term might cause the traditional PPN implementation of the SEP/GNEP to fail, albeit in practice one would not expect any significant violation to arise.} (considered as a fundamental constant of Nature, but with its value left unspecified), \eq{\de \Sigma = \de^{m-1} x \, \sqrt{\abs{h}}} is the volume element on the \eq{\ton{m-1}}-dimensional boundary \eq{\pd \Scr{U}} (\eq{h} is the determinant of the induced metric), and \eq{K} is the trace of the extrinsic curvature.  We notice that the presence of the boundary term is strictly necessary to make the variational problem well posed, as it permits for an identical cancellation of another boundary term containing \eq{\nabla_{\! c} \delta\matg_{ab}} that appears when the metric is varied by \eq{\delta \matg_{ab}}.  Then, to get Einstein's field equations, all one has to set is \eq{\delta \matg_{ab} = 0} at the boundary~\cite{Wald:1984rg, Dyer:2008hb}.  This leads to the following expression for the tensor \eq{E_{ab}}:
\eqd{E_{ab}=\frac{1}{8 \pi G} \ton{G_{ab} + \Lambda \, \matg_{ab}}.}

From the results of Sec.~\ref{Ss:var}, it is obvious that this theory, being purely metric, passes our test of the GWEP.  However, the validity of condition~\eqref{eq:cons} can also be checked independently, via a full calculation of \eq{\Cal{E}_{ab}} and its covariant divergence with respect to the background metric.  Since, at first order,
\eqd{\Cal{E}_{ab} = \frac{1}{8 \pi G} \ton{\Cal{G}_{ab} + \Lambda \gamma_{ab}} \,, \label{eq:cosmoconst}}
where \eq{\epsilon \, \Cal{G}_{ab}} is the first-order term in the expansion of the Einstein tensor (see Eq.~\eqref{G1} for the full-length expression), condition~\eqref{eq:cons} reduces simply to 
\eqd{\bar{\nabla}^b \Cal{G}_{ab} = - \Lambda \bar{\nabla}^b \gamma_{ab} 
\label{porcoqua}\;.}
A somewhat lengthy, but straightforward calculation (see Appendix~\ref{App:B}) shows that the quantity \eq{\bar{\nabla}^b \Cal{G}_{ab}} is given by Eq.~\eqref{divG1}.  However, as stated above, we work on a background which is a solution of the vacuum field equations, so \eq{\bar{R}_{ab} = \Lambda \, \bar{\matg}_{ab}}.  Plugging this into the general expression~\eqref{divG1}, and using the metric-compatibility condition \eq{\bar{\nabla}_{\! c} \, \bar{\matg}_{ab} = 0}, we find that Eq.~\eqref{porcoqua} is indeed satisfied.

This assures that, in Einstein's theory with an arbitrary cosmological constant (hence, in particular, also for \eq{\Lambda = 0}), the GWEP holds true, and the world lines of bodies with negligible size but non-negligible self-gravitation in an empty background are geodesics.

%--------------------------------
\subsection{Scalar-tensor theories}
\label{ss:scalartens}
%--------------------------------

Let us now apply our method to a prototypical example of an extended theory of gravity; that is, one going beyond the framework of general relativity.  In this particular case, it is already known that the theory does not obey the GNEP.  We consider the Brans--Dicke theory~\cite{faraoni:2004cos}, which corresponds to the gravitational action 
\eqd{A_{\text{BD}} = \frac{1}{16 \pi} \int_{\mathscr{U}} \de \Omega \ton{\phi R - \frac{\omega}{\phi} \, \matg^{ab} \, \nabla_{\! a} \phi\, \nabla_{\! b} \phi} + \text{b.t.} \,,
\label{eq:bdgravaction}}
where \eq{\omega} is a nonzero constant and, here and in the following, ``b.t.'' stands for an appropriate boundary term.  The main feature of this theory is the presence of an additional gravitational degree of freedom, in the form of a single scalar field \eq{\phi}.  A comparison with the action~\eqref{eq:einhilact} for general relativity shows that we have simply replaced the constant \eq{{1}/{G}} with a function \eq{\phi}, and then given such field a dynamical behaviour via the additional kinetic term.  Thus, we have formally promoted Newton's constant \eq{G} to a function on spacetime.   As in the case of Einstein's theory, the boundary term makes the variational formulation well posed.

By varying \eq{A_{\text{BD}}} above with respect to \eq{\matg^{ab}}, one gets field equations of the form~\eqref{eq:Gen-eq}, with 
\speq{\! E_{ab} = &\frac{1}{8\pi} \left(\phi \, G_{ab} - \frac{\omega}{\phi} \left(\nabla_{\! a} \phi \, \nabla_{\! b} \phi - \frac{1}{2} \, \matg_{ab}\nabla^c \phi \, \nabla_{\! c} \phi \right)\right. \\
&\quad- \nabla_{\! a} \nabla_{\! b} \phi + \matg_{ab} \square \phi \bigg) \,, \label{eq:bransdicke}}
where \eq{\square := \matg^{ab} \nabla_{\! a} \nabla_{\! b}} is the d'Alembertian operator. In addition, there is another scalar equation accounting for the dynamics of the long-range field \eq{\phi}, namely
\subeq{\eqd{R - \frac{\omega}{\phi^2} \, \nabla^{a} \phi \, \nabla_{\! a} \phi + \frac{2 \omega}{\phi} \, \square \phi = 0 \,. \label{eq:bdscalar}}
This, together with Eq.~\eqref{eq:bransdicke}, yields
\eqd{\square \phi = \frac{8 \pi}{3 + 2 \omega} \, T \,, \label{urcaurca}}}
where \eq{T = \matg^{ab} T_{ab}}.  Equations~\eqref{eq:bdscalar} and~\eqref{urcaurca} can be used interchangeably.

Since this theory contains the scalar field \eq{\phi} in its gravitational sector, it is not purely metric and violates condition~\eqref{i)}.  Hence, condition~\eqref{eq:cons} is also violated, and the GWEP does not hold.  It is instructive, however, to consider this result from another point of view.  Let us pick a background vacuum solution, so \eq{\bar{T}_{ab} = 0} and condition~\eqref{ii)} is satisfied.  The zeroth-order field equation \eq{\bar{E}_{ab} = 0} can be rearranged as \eq{\bar{G}_{ab} = \bar{T}^{(\phi)}_{ab}},  where 
\speq{\bar{T}^{(\phi)}_{ab} = &\frac{\omega}{\bar{\phi}^2} \ton{\bar{\nabla}_{\! a} \bar{\phi} \, \bar{\nabla}_{\! b} \bar{\phi} - \frac{1}{2} \, \bar{\matg}_{ab} \, \bar{\nabla}^{c} \bar{\phi} \, \bar{\nabla}_{\! c} \bar{\phi}} \\
&\quad+ \frac{\bar{\nabla}_{\! a} \bar{\nabla}_{\! b} \bar{\phi}}{\bar{\phi}} - \bar{\matg}_{ab} \, \frac{\bar{\square} \bar{\phi}}{\bar{\phi}}\;,
\label{eq:einstbd}}
with the indices raised with, and the d'Alembertian operator built out of, the background metric alone.  Thus, \eq{\bar{T}^{(\phi)}_{ab}} can be thought of as a ``matter'' stress-energy-momentum tensor associated with the dynamical long-range field \eq{\phi} at zeroth order.  By such redefinition, we see immediately that we fall back into something which is, formally, just general relativity with a particular non-empty background --- hence the failure of the GWEP.

We can summarise the results as follows: for Brans--Dicke gravity, the existence of an additional gravitational degree of freedom besides the metric, i.e., the non-purely metric character of the theory, leads to the failure of the GWEP because of the non-vanishing first term in Eq.~\eqref{popopoff}. Alternatively, upon rearranging the field equations so as to have some sort of ``pure general relativity plus exotic scalar matter'', the other terms in Eq.~\eqref{popopoff} spoil the equivalence, and once again certify the failure of the geodesic motion for self-gravitating test bodies.

A straightforward calculation of \eq{ \bar{\nabla}^b \Cal{E}_{ab}}, as done in the case of general relativity, would be very lengthy and cumbersome.  Just to convey an idea of the sort of objects involved, and of the power of the general approach developed in Sec.~\ref{Ss:var}, we exhibit the explicit form of the first-order tensor \eq{\Cal{E}_{ab}}, which reads
\begin{widetext}
\speq{\Cal{E}_{ab} = &\frac{1}{8\pi}\left(\bar{\phi}\; \Cal{G}_{ab} + \chi\, \bar{G}_{ab} - \frac{\omega}{\bar{\phi}} \ton{\bar{\nabla}_{\! a} \bar{\phi}\, \bar{\nabla}_{\! b} \chi + \bar{\nabla}_{\! a} \chi\, \bar{\nabla}_{\! b} \bar{\phi}} + \frac{\omega\,\chi}{\bar{\phi}^2}\, \bar{\nabla}_{\! a} \bar{\phi}\, \bar{\nabla}_{\! b} \bar{\phi} - \frac{\omega}{2 \bar{\phi}}\, \bar{\matg}_{a b}\, \gamma^{c d} \bar{\nabla}_{\! c} \bar{\phi}\, \bar{\nabla}_{\! d} \bar{\phi} \right. \\
&+ \frac{\omega}{2 \bar{\phi}}\, \gamma_{a b}\,\bar{\matg}^{cd}\, \bar{\nabla}_{\! c} \bar{\phi}\, \bar{\nabla}_{\! d} \bar{\phi} + \frac{\omega}{\bar{\phi}}\, \bar{\matg}_{a b}\,\bar{\matg}^{cd}\, \bar{\nabla}_{\! c} \bar{\phi}\, \bar{\nabla}_{\! d} \chi -\frac{\omega\,\chi}{2\bar{\phi}^2}\,\bar{\matg}_{ab}\,\bar{\matg}^{cd}\,\bar{\nabla}_{\! c}\bar{\phi}\,\bar{\nabla}_{\! d}\bar{\phi} + {\Xi^c}_{ab} \bar{\nabla}_{\! c} \bar{\phi} \\
&- \biggl.\bar{\matg}_{a b} \gamma^{cd}\, \bar{\nabla}_{\! c} \bar{\nabla}_{\! d} \bar{\phi} - \bar{\matg}_{a b}\, \bar{\matg}^{cd}\, {\Xi^e}_{cd} \bar{\nabla}_{\! e} \bar{\phi} + \bar{\matg}_{a b}\,\bar{\matg}^{cd}\, \bar{\nabla}_{\! c}\bar{\nabla}_{\! d} \chi + \gamma_{ab}\,\bar{\matg}^{cd}\, \bar{\nabla}_{\! c}\bar{\nabla}_{\! d} \bar{\phi} - \bar{\nabla}_{\! a} \bar{\nabla}_{\! b} \chi \biggr),}
\end{widetext}
where \eq{{\Xi^{a}}_{bc}} is given by Eq.~\eqref{Xi!}, \eq{\Cal{G}_{ab}} is given by Eq.~\eqref{eq:calligrG}, and \eq{\bar{G}_{ab}} is related to \eq{\bar{\phi}} and its derivatives through the background field equation \eq{\bar{G}_{ab} = \bar{T}^{(\phi)}_{ab}}, with $\bar{T}^{(\phi)}_{ab}$ given by  Eq.~\eqref{eq:einstbd}.

Upon generalising the action~\eqref{eq:bdgravaction} to the one of an arbitrary scalar-tensor theory, namely
\eqd{A_{\text{ST}} = \frac{1}{16\pi} \int_{\mathscr{U}} \!\! \de \Omega \ton{\phi R - \frac{\omega \ton{\phi}}{\phi} \nabla^{a} \phi \nabla_{\! a} \phi +  V \ton{\phi}} + \text{b.t.},}
the conclusion about the failure of the GWEP clearly does not change, as the theory is still not purely metric.  By the same token, one might say that, upon rewriting the field equations as ``general relativity with scalar matter'', the generic zeroth-order tensor \eq{\bar{T}^{(\phi)}_{ab}} would still prevent the validation of the GWEP.

Noticeably, the same logic can be applied to the much wider class of multi-scalar-tensor theories~\cite{faraoni:2004cos}. Either the appearance of more than one additional scalar degree of freedom in~\eqref{popopoff} (where the first term ought to be traded for a sum over the number of scalar fields available), or the presence of a cumulative nonzero background stress-energy tensor for the scalar fields --- also in matter vacuum --- will lead to the failure of condition~\eqref{eq:cons}.

%--------------------------------
\subsection{Higher-order theories and their subtleties}
\label{ss:extended}
%--------------------------------

The last paragraphs above allow us to discuss yet another interesting case, viz., a vast sub-class of the so-called higher-order theories of gravity, for which the true status of the GWEP is still somewhat debated~\cite{will}.  As we shall see, while the core of the method used so far remains unchanged, there are some crucial niceties worth pointing out, to shine a light on some misconceptions afflicting the zoology of higher-order actions.

We begin with the basics. In the simplest cases of higher-curvature theories, the action is built substituting the scalar curvature \eq{R} in Eq.~\eqref{eq:einhilact}, with a more general arbitrary function \eq{f \! \ton{R}}, supposed to be analytic in its argument. The resulting field equations --- if existing at all; see below --- generally exhibit derivatives of order higher than \eq{2}, which can be sometimes reformulated in terms of excitations of additional modes of the graviton propagator and other physically measurable effects~\cite{Sotiriou:2008rp}. The next step is to identify even further scalar combinations of \eq{?R_abc^d?}, \eq{R_{ab}}, \eq{R}, and functions thereof, exhausting the landscape of higher-curvature theories.

At a first glance, it might appear that the only gravitational degrees of freedom are those encoded in the metric, so that e.g., a generic \eq{f \! \ton{R}} theory is purely metric, like general relativity. Hence, the entire class of theories should satisfy the GWEP.  The same argument might then be repeated identically for more general cases.

This conclusion is actually wrong.  The reason, however, is partially concealed by the way the gravitational actions are built, and only a deeper inspection allows for a full understanding of the pitfall. The key element is the role played by the boundary terms in the action.

Let us start with the exemplary case of an \eq{f \! \ton{R}} theory. The action can be written as
\eqd{A_{f\! \ton{R}} = \frac{1}{16 \pi} \int_{\mathscr{U}} \de \Omega \; f \! \ton{R} 
+ \text{b.t.} \,, \label{eq:actionfR}}
where the boundary term basically mirrors that of the Einstein--Hilbert action~\cite{Guarnizo:2010xr}.  Upon performing the variation, the resulting field equations read
\speq{f' \! \ton{R} R_{a b} - &\frac{1}{2}\,f \! \ton{R} \matg_{a b} - \nabla_{\! a} \nabla_{\! b} f' \! \ton{R} \\
&+ \matg_{a b}\, \square f' \! \ton{R} = 8\pi\, T_{ab}\;, \label{eq:fieldeqfR}}
where \eq{f' \! \ton{R} := \de f \! \ton{R} / \de R}. To get this result, however, one is forced to impose the condition \eq{\delta R = 0} on the boundary~\cite{Guarnizo:2010xr}, which amounts to fixing the value of a combination of the metric and its first and second derivatives.  This seems too strong a requirement, even for a fourth-order theory as this one.  It thus seems that the functional derivative with respect to \eq{\matg^{ab}} of the action~\eqref{eq:actionfR} is not defined, and that the field equations~\eqref{eq:fieldeqfR} cannot be derived by a variational principle, after all.

The way out of this difficulty is to recognise that, in fact, there is another degree of freedom, associated with a scalar field \eq{\phi}, which is  kinematically independent of \eq{R} but dynamically related to it in an algebraic manner, so that \eq{\delta R = 0} is not a condition on the second derivatives of the metric, but rather on \eq{\phi}.  The action~\eqref{eq:actionfR} can be equivalently rewritten as that of a particular scalar-tensor theory where one of the field equations is just \eq{\phi := f' \! \ton{R}} \cite{Sotiriou:2007zu}.  Only in this new formulation can the boundary term be introduced and all problems removed.\footnote{Already at the level of elementary, one-dimensional mechanical systems, it is possible to convince oneself that the lack of a proper boundary term, or the need to impose too many conditions at the boundary, can easily lead to any sort of inconsistencies, loop arguments, and ill-posedness of the equations of motion~\cite{Dyer:2008hb}.}  The variational principle becomes then well posed and our treatment in Sec.~\ref{Ss:var} applies.  Thus, the seemingly purely metric \eq{f \! \ton{R}}\ theories of gravity unveil themselves as generalised scalar-tensor theories in disguise~\cite{Sotiriou:2008rp}, and do not pass our test for the GWEP.

This scheme can be generalised. All known cases of theories with higher-order field equations require some additional constraints on the boundary (the ``higher momenta'') to get the field equations from the action.  Such extra conditions go beyond the standard ones expected from a purely metric framework, and could be avoided only introducing  boundary integrals analogous to the Gibbons--Hawking--York counter-term of general relativity.  No such boundary terms, however, have been found yet for such theories.  At the same time, theories of the kind \eq{f\! \ton{R , \square R , \ldots}} --- which indeed produce higher-order field equations --- have been suitably reformulated into equivalent multi-scalar-tensor theories with second-order field equations only~\cite{Schmidt:1989zz,Gottlober:1989ww,Baykal:2013gfa}.  The variational formulation of these equivalent theories is well posed, but the presence of additional degrees of freedom implies that the GWEP is violated.

These results suggest that a theory with higher curvatures in the action, but only second-order field equations, may provide a different answer, and enlarge the category of GWEP-validating frameworks.  The question might seem bizarre, as \eq{A_{\text{GR}}} given by Eq.~\eqref{eq:einhilact} is the only non-trivial action one can write in four dimensions, which gives second-order field equations (while being also linear in the curvature).  This issue, however, is not so trivial in higher dimensions.  It is known since 1971 that there is a whole class of higher-curvature theories, described by the so-called Lanczos--Lovelock actions~\cite{Lovelock:1971yv}, leading to at most second-order field equations.  For example, in five dimensions, besides the 5D-version of general relativity, one has also Gau\ss{}--Bonnet gravity~\cite{Padmanabhan:2010zzb}, which has an action characterised by terms quadratic in the curvature.  This theory (and all its higher-dimensional companions) is therefore an optimal test bench for discriminating another aspect that might be crucial in the realisation or failure of the GWEP.

%--------------------------------
\subsection{Gau\ss{}--Bonnet and Lanczos--Lovelock gravity}
\label{ss:gaussbonnet}
%--------------------------------

The general Lanczos--Lovelock gravitational action in an \eq{m}-dimensional spacetime has the compact form~\cite{Padmanabhan:2010zzb}
\eqd{A_{\text{LL}} =\frac{1}{16\pi G}  \int_{\mathscr{U}} \de \Omega \;  Q_{abcd}\, R^{abcd} + \text{b.t.} \,. \label{eq:GB-action}}
The tensor \eq{Q_{abcd}} retains all the algebraic symmetries of the Riemann tensor, and has vanishing divergence, \eq{\nabla_{\! d} ?Q_abc^d? = 0}.  Such object is designed so as to provide, for each value of \eq{m}, all the combinations giving rise to second-order field equations, but not higher derivative terms. This feature, together with variational arguments, makes it possible to recognise the tensor \eq{Q_{abcd}} as a polynomial in the curvature tensor with constant coefficients, whose degree depends on the dimension of spacetime.  The zeroth-order term of such polynomial is simply the combination \eq{\matg_{a [ c}\, \matg_{d ] b}}.  When contracted with the Riemann curvature tensor, this gives the scalar curvature \eq{R}, thus leading to the Einstein--Hilbert action~\eqref{eq:einhilact}.

The search for the boundary terms is in this case particularly labourious.  Yet, such terms exist~\cite{bunch, Sendouda:2011hq, Myers:1987yn, Mukhopadhyay:2006vu}, and make the variational problem well posed for all Lanczos theories.  As a consequence, and this is of course relevant for our purposes, Lovelock gravity is a purely metric theory.\footnote{There has been a recent attempt~\cite{Brustein:2012uu} to describe Lanczos--Lovelock actions in terms of pure \eq{m}-dimensional general relativity, plus a convenient number of \eq{3}-form fields and their covariant derivatives.  Yet, an accurate inspection of Ref.~\cite{Brustein:2012uu} --- particularly of Eq.~(21) therein --- shows that the new fields supposedly encoding the additional gravitational degrees of freedom have identically vanishing variation, which makes them just auxiliary variables, rather than truly dynamical ones.} 

The lowest nontrivial element of the Lanczos--Lovelock class is the Gau\ss{}--Bonnet theory --- emerging from \eq{m = 5} onwards --- for which it is
\speq{Q_{abcd} = \alpha &\left( R_{abcd} + G_{bc} \matg_{ad} - G_{ac} \matg_{bd} \right. \\
&+\left. R_{ad} \matg_{bc} - R_{bd} \matg_{ac} \right) \,,}
where \eq{\alpha} is a constant with dimensions \eq{\ton{\text{length}}^2}. The corresponding Lagrangian density is 
\eqd{Q_{abcd} R^{abcd} = \alpha \ton{R^2 - 4 R^{ab} R_{ab} + R^{abcd} R_{abcd}} \,. \label{eq:GBlagrangian}}
When one also adds the zeroth-order term \eq{R}, the variation of the action with respect to the metric gives 
\eqd{E_{ab}=\frac{1}{8\pi G}\,\left(G_{ab} + \alpha \, Y_{ab} \right), 
\label{eq:gbfieldeq}}
where the tensor \eq{Y_{ab}} is
\speq{Y_{ab} := \, &2 R R_{ab} - 4 ?R_ac? ?R^c_b? - 4 ?R^cd? ?R_acbd? + 2 ?R_a^cde? ?R_bcde?\\
&- \frac{1}{2}\, \matg_{ab} \left( R^2 - 4 ?R^cd? ?R_cd? + ?R^cdef? ?R_cdef? \right) \,. \label{eq:GB-field}}
An analogous recipe permits to build all the subsequent Lanczos--Lovelock terms compatible with the given spacetime dimension.  The higher-curvature corrections to the Einstein--Hilbert action are all tailored so that no field equations of order higher than \eq{2} can emerge.

Since, as noticed, any Lanczos--Lovelock theory is purely metric, condition~\eqref{i)} is satisfied, hence \eq{\bar{\nabla}^b \Cal{E}_{ab} = 0} in vacuum.  Therefore, the GWEP is valid within the entire class of such theories.  This last result has been advanced in Ref.~\cite{Deruelle:2003ps} for the Gau\ss{}--Bonnet theory, as a consequence of a completely different point of view --- namely, that of Noether-like conserved currents.

There are some interesting consequences. First, we have now another class of theories fulfilling the requirements to obey the GWEP.  Such theories complement general relativity (actually, its \eq{m}-dimensional version, for each \eq{m \ge 4}), which was already known to abide by the principle~\cite{deruelle, Deruelle:2003ps}.  

Furthermore, and this is perhaps more interesting, for \eq{m} strictly greater than \eq{4}, there are several competing purely metric theories of gravity, all providing an almost-geodesic motion for self-gravitating test bodies.  Hence, there might be other criteria than the GWEP, playing a role in further selecting a subset within this class.  Presumably, they will have to deal  with the specific form of gravitational self-coupling.

In this sense, once again Lanczos--Lovelock theories prove to be ``special'' within the landscape of extended theories of gravitation, and a ``natural'' generalisation of Einstein's seminal model~\cite{Mukhopadhyay:2006vu, Padmanabhan:2010zzb}. At the same time, the difference in the actual construction of the gravitational action, and the resulting different phenomenology~\cite{Choquet-Bruhat:2009xil, Golod:2012yt}, seem to indicate that other criteria may possibly be introduced, to build an even finer taxonomic sieve.

%===========================
\section{Concluding remarks}
\label{S:7}
%===========================

In this work, we have shown how it is possible to find novel and useful bits of information in the seemingly outdated principle of equivalence, and we have built a formal and operative protocol to single out purely metric theories of gravitation starting from heuristic concepts and elementary considerations about the behaviour of self-gravitating test bodies.

Notwithstanding the inevitably perturbation-based scheme, and the number of hypotheses and restrictions necessary for the formulation to be meaningful, we deem to have finally given the gravitational weak equivalence principle a truly constructive role, similar to that of other equivalence principles such as the WEP and EEP, in the selection of gravitation theories.

The technique proposed overcomes at once the difficulties and limitations intrinsic to the PPN formalism (and similar methods developed for solar system experiments or for astronomical conditions alike), relying on a much more immediate inspection of the first-order action for a given theory of gravity, provided that such action is equipped with boundary terms that make the variational principle well posed.  It can be adapted seamlessly to spacetimes of arbitrary dimension, thus permitting a quick access to all those theories for which the PPN formalism and similar procedures have not yet been developed.

Our conclusions fully agree with the common interpretation of the strong equivalence principle (or rather, with a restricted version of it), in the sense that only within purely metric theories of gravity does universality of free fall extend to self-gravitating bodies~\cite{will}, and only if the background environment is freed from matter content other than the body under consideration~\cite{Mino}.  However, we provide a formally sound understanding of the roots of these generally accepted notions.  Our method confirms or disproves the GWEP in any case where other independent techniques provide an answer. This has been shown working out explicitly the examples of general relativity (with or without a cosmological constant term), the Brans--Dicke theory, and general (multi-)scalar-tensor theories.  But we have also presented new results, concerning higher-order and Lanczos--Lovelock theories.

Since the GWEP is ultimately satisfied only by purely metric theories, testing it becomes a matter of acknowledging the existence of extra gravitational degrees of freedom.  The latter are not always evident, but can be  hidden in the structure of the action (as it happens for the higher-order theories discussed in Sec.~\ref{ss:extended}), eventually cropping up through the impossibility of adding proper boundary terms that make the functional derivative with respect to the metric well defined.  On the other hand, a failure of the GWEP may be considered a further hint that such hidden variables are present, and that only a rethinking of the theories in terms of more degrees of freedom will lead to their variational formulation becoming well posed.  Thus, not only is the GWEP singling out purely metric theories; it can also become an independent tool to unveil the deep structure of gravity theories, which is sometimes concealed by a particular choice of variables.

An issue that remains untouched is the relation between our findings and the recent proposal in Refs.~\cite{Gerard:2006ia,Gerard:2008nc}.  It is difficult, at this stage, to say whether some form of general agreement could be reached, because the results of the two approaches can be compared only in the cases of general relativity and scalar-tensor theories (where they agree, of course).  Indeed, the  extension of condition~\eqref{eq:gerard} to more general scenarios is not provided in the original references, and might well require adjustments when passing, e.g., to the context of Lanczos--Lovelock gravity.\footnote{Taken as is, Eq.~\eqref{eq:gerard} is not verified already at the level of a single black hole-like solution for Gau\ss{}--Bonnet theory, let alone for any vacuum solution. This fact, together with the formally different version of the same condition provided for scalar-tensor theories, viz. \eq{\nabla_{\! d} (\phi ?R_abc^d?) = 0}, strongly suggests that~\eqref{eq:gerard} may not be a universal law expressing the SEP/GWEP in a gauge-inspired representation of any extended theory of gravity.}

As an aside, since in higher-dimensional spacetimes it has been found that not only general relativity, but also the whole class of higher-curvature but not higher-derivative Lanczos--Lovelock theories comply with the GWEP, there is potential room for improvement in the selection rule, perchance integrating our test with some restrictions on the kind of self-interaction of the gravitational degrees of freedom. In this respect, it is worth noticing that the \eq{m}-dimensional version of Einstein's theory contains the minimal self-coupling conceivable in a non-linear context, whereas the complexity increases with the order of the Lanczos--Lovelock polynomials. It is far from clear, though, whether and how this might be implemented thoroughly as an additional selection criterion.

Our work has attempted at repeating, on a much smaller scale, the historical process through which a set of heuristic statements was turned into a  formal synthesis able to motivate precision-test experiments. In doing so, we have further refined our understanding of some of the relations among the various extended theories of gravity, and partially satiated a ``taxonomic hunger'' which is germane to the community of researchers~\cite{Sotiriou:2007zu}. Far from being a dry exercise in abstract systematics, this urge to detail the family tree of gravitational theories is motivated by both puzzling observational evidence of an unexplained behaviour of our universe~\cite{Ade:2013zuv}, and daring theoretical proposals (e.g., the AdS/CFT duality). And it is somewhat comforting to see that the most fundamental, although perhaps slightly na{\"i}ve, intuitions of our predecessors can provide, still nowadays, robust guidelines for a better understanding of gravity.

\begin{acknowledgments}
E.D.C. gratefully acknowledges: Vincenzo Vitagliano, Goffredo Chirco, Iwona Mochol, and Arletta Nowodworska, for their inspiring support; Eric Poisson, Ian Vega, Lorenzo Sindoni, Alessio Belenchia, Daniele Vernieri and Noemi Frusciante, for their keen remarks and stimulating questions; and Antonio Romano, for illuminating discussions. S.L. thanks Thanu Padmanabhan for insightful suggestions and shrewd considerations.
\end{acknowledgments}

%-------------------------------------------------
\appendix
%-------------------------------------------------

%===========================
\section{First-order perturbations}
\label{App:A}
%===========================

In this Appendix, we collect a number of useful expressions for the differences, to the first order in \eq{\epsilon}, between the geometric objects built out of two metrics \eq{\matg_{ab}} and \eq{\bar{\matg}_{ab}} connected via the relation~\eqref{eq:metricexp}.  Thus, all the equations presented here hold only up to order \eq{\epsilon}.  First of all, we note that Eq.~\eqref{eq:metricexp} implies 
\eqd{\matg^{ab} = \bar{\matg}^{ab} - \epsilon \, \gamma^{ab} \,, \label{invg}}
where \eq{\bar{\matg}^{ab}} is the inverse of \eq{\bar{\matg}_{ab}}, and \eq{\gamma^{ab} := \bar{\matg}^{ac} \, \bar{\matg}^{bd} \, \gamma_{cd}}.  

To find the relation between the determinants of the metric coefficients, let us first expand \eq{\matg} around the unperturbed metric \eq{\bar{\matg}_{ab}}:
\eqd{\matg = \bar{\matg} + \epsilon \, \diffp{\matg}{{{\matg_{ab}}}} \, \gamma_{ab} \;,}
where the partial derivatives are evaluated at \eq{\matg_{ab} = \bar{\matg}_{ab}}.  
Using the property \eq{{\partial \matg}/{\partial \matg_{ab}} = \matg \, \matg^{ab}}, and defining \eq{\gamma := \bar{\matg}^{ab} \, \gamma_{ab}}, we find the simple relation
\eqd{\matg = \bar{\matg} \ton{1 + \epsilon \, \gamma} \,. \label{ggbar}}

The Christoffel symbols \eq{{\Gamma^a}_{bc}} and \eq{?{{{\bar{\Gamma}}}}^a_bc?} of the metrics \eq{\matg_{ab}} and \eq{\bar{\matg}_{ab}}, respectively, are related as
\eqd{{\Gamma^a}_{bc} = ?{{{\bar{\Gamma}}}}^a_bc? + \epsilon \, {\Xi^a}_{bc}\;, \label{Christ}}
where \eq{{\Xi^a}_{bc}:=\bar{\matg}^{ad} \, \Xi_{dbc}}, and the tensor \eq{\Xi_{abc}} is
\eqd{\Xi_{abc} = \frac{1}{2} \, \left(\bar{\nabla}_{\! b} \gamma_{ca} + \bar{\nabla}_{\! c} \gamma_{ba} - \bar{\nabla}_{\! a} \gamma_{bc}\right). \label{Xi!}}
This result can be easily obtained using the expression 
\speq{\nabla_{\! a} \matg_{bc} &= \bar{\nabla}_{\! a} \matg_{bc} - \epsilon \, {\Xi^d}_{ab} \, \matg_{dc} - \epsilon \, {\Xi^d}_{ac} \, \matg_{bd} \\
&= \bar{\nabla}_{\! a} \bar{\matg}_{bc} + \epsilon \, \bar{\nabla}_{\! a} \gamma_{bc} - \epsilon \, \Xi_{cab} - \epsilon \, \Xi_{bac} \,,}
which follows from Eq.~\eqref{Christ}.  Since the covariant derivatives \eq{\nabla_{\! a}} and \eq{\bar{\nabla}_{\! a}} are associated with \eq{\matg_{ab}} and \eq{\bar{\matg}_{ab}}, respectively, the compatibility condition for the Riemannian connection gives \eq{\nabla_{\! a} \matg_{bc} = \bar{\nabla}_{\! a} \bar{\matg}_{bc} = 0}.  Thus, 
\eqd{\Xi_{cab} + \Xi_{bac} = \bar{\nabla}_{\! a} \gamma_{bc} \,.}
Equation~\eqref{Xi!} is then obtained following the same steps by which one finds the usual expression for the Christoffel symbols in terms of partial derivatives of the metric.

The first-order difference \eq{\epsilon \, {\Cal{R}_{abc}}^d = {R_{abc}}^d - ?{{{\bar{R}}}}_abc^d?} between the Riemann curvature tensors follows from Eq.~\eqref{Christ}, and one has
\eqd{{\Cal{R}_{abc}}^d = \bar{\nabla}_{\! b} {\Xi^d}_{ac} - \bar{\nabla}_{\! a} {\Xi^d}_{bc} \,.}
This implies, for the difference \eq{\epsilon \, \Cal{R}_{ab} = R_{ab} - \bar{R}_{ab}} between the Ricci tensors,
\eqd{\Cal{R}_{ab} = {\Cal{R}_{acb}}^c = \bar{\nabla}_{\! c} {\Xi^c}_{ab} - \bar{\nabla}_{\! a} {\Xi^c}_{cb} \,; \label{Riccissimo}}
and, for the difference \eq{\epsilon \, \Cal{R} = R - \bar{R}} between the curvature scalars \eq{R = \matg^{ab} R_{ab}} and \eq{\bar{R} = \bar{\matg}^{ab} \bar{R}_{ab}},
\eqd{\Cal{R} = \bar{\matg}^{ab} \, \bar{\nabla}_{\! c} {\Xi^c}_{ab} - \bar{\matg}^{ab} \, \bar{\nabla}_{\! a} {\Xi^c}_{cb} - \gamma^{ab} \bar{R}_{ab} \,,}
where Eqs.~\eqref{invg} and~\eqref{Riccissimo} have been used.  

Finally, for the difference \eq{\epsilon \, \Cal{G}_{ab} = G_{ab} - \bar{G}_{ab}} between the Einstein tensors we find, defining \eq{{\Xi^{ab}}_{b} := \bar{\matg}^{bc} \, {\Xi^a}_{bc}}:
\speq{\Cal{G}_{ab} &= \Cal{R}_{ab} - \frac{1}{2} \, \bar{\matg}_{ab} \, \Cal{R} - \frac{1}{2} \, \bar{R} \, \gamma_{ab} = \bar{\nabla}_{\! c} {\Xi^c}_{ab} - \bar{\nabla}_{\! a} {\Xi^c}_{cb} \\
&- \frac{\bar{\matg}_{ab}}{2} \, \bar{\nabla}_{c} {\Xi^{cd}}_{d} + \frac{\bar{\matg}_{ab}}{2} \bar{\nabla}^{c} {\Xi^e}_{ec} + \frac{\bar{\matg}_{ab}}{2} \gamma^{cd} \bar{R}_{cd} - \frac{\gamma_{ab}}{2} \bar{R} \,. \label{G1}}
%

%===========================
\section{Calculation of $\bar{\nabla}^b \Cal{G}_{ab}$}
\label{App:B}
%===========================

The quantity \eq{\bar{\nabla}^b \Cal{G}_{ab}} intervenes frequently in the calculation of \eq{\bar{\nabla}^b \Cal{E}_{ab}} --- noticeably, in Sec.~\ref{ss:GR} ---, so we evaluate it here in full generality.  We begin by substituting the expression~\eqref{Xi!} for \eq{{\Xi^a}_{bc}} into Eq.~\eqref{G1}, to obtain
\speq{\Cal{G}_{a b} = &\frac{1}{2} \, \Big[\ton{\bar{\nabla}^c \bar{\nabla}_{\! a} \gamma_{b c} + \bar{\nabla}^c \bar{\nabla}_{\! b} \gamma_{a c}} - \bar{\nabla}^c \bar{\nabla}_{\! c} \gamma_{a b} - \bar{\nabla}_{\! a} \bar{\nabla}_{\! b} \gamma \\
&- \bar{\matg}_{a b} \ton{\bar{\nabla}^c \bar{\nabla}^d \gamma_{c d} - \bar{\nabla}^c \bar{\nabla}_{\! c} \gamma} + \bar{\matg}_{ab} \gamma^{cd} \bar{R}_{cd} - \gamma_{ab} \, \bar{R}\,\Big]. \label{eq:calligrG}}
In a flat background spacetime (a situation common, for instance, in the study of gravitational radiation~\cite{Wald:1984rg}), it is a straightforward exercise to show that \eq{\pd^b \Cal{G}_{ab} = 0}, the key point in the proof being a heavy use of the commutative property for partial derivatives.  In the case of a non-flat background, on the other hand, switching covariant derivative operators \eq{\bar{\nabla}_{\! a}} generates instances of the Riemann and Ricci tensors.  The three terms in \eq{\bar{\nabla}^b \Cal{G}_{ab}} where this happens can be written, rearranging the indices and using the property \eq{\bar{\nabla}_{\! a} \bar{\matg}_{cd} = 0}, as:
\gat{\begin{split}\bar{\nabla}_{\! b} &\bar{\nabla}_{\! c} \bar{\nabla}_{\! a} \gamma^{bc} - \bar{\nabla}_{\! a} \bar{\nabla}_{\! b} \bar{\nabla}_{\! c} \gamma^{bc} = \bar{R}_{abcd} \, \bar{\nabla}^d \gamma^{bc} \\
&+2 \bar{R}_{ab} \bar{\nabla}_{\! c} \gamma^{bc} + 2 \bar{\nabla}_{\! b} \bar{R}_{ac} \gamma^{bc} - \bar{\nabla}_{\! a} \bar{R}_{bc} \gamma^{bc} \,; \label{casin1} \end{split} \\[1em]
\bar{\nabla}_{\! b} \bar{\nabla}_{\! c} \bar{\nabla}^b {\gamma_a}^c-\bar{\nabla}_{\! c}\bar{\nabla}_{\! b}\bar{\nabla}^b{\gamma_a}^c = - \bar{R}_{bcda} \, \bar{\nabla}^b \gamma^{cd} \,; \label{casin2} \\[1em]
\bar{\nabla}_{\! a}\bar{\nabla}_{\! b}\bar{\nabla}^b\gamma-\bar{\nabla}_{\! b}\bar{\nabla}_{\! a}\bar{\nabla}^b\gamma=-\bar{R}_{ab}\bar{\nabla}^b\gamma\;;
\label{casin3}}
where in Eq.~\eqref{casin1} we have used the general identity~\eqref{eq:rotR} for the background quantities.  Using these expressions, we find at the end
\speq{\bar{\nabla}^b \Cal{G}_{ab} = &\frac{1}{2} \left( 2 \bar{R}_{ab} \bar{\nabla}_{\! c}\gamma^{bc} + 2 \gamma^{bc} \bar{\nabla}_{\! b}\bar{R}_{ac} - \bar{R}_{ab} \bar{\nabla}^b \gamma \right. \\[0.5em]
&\left. + \bar{R}_{bc} \bar{\nabla}_{\! a} \gamma^{bc} - \bar{R} \bar{\nabla}^b \gamma_{ab} - \gamma_{ab} \bar{\nabla}^b \bar{R} \right).
\label{divG1}}
%

%===========================
%
%
%
\end{document}